# A Natural Language-Inspired Multi-label Video Streaming Traffic Classification Method Based on Deep Neural Networks

*Yan Shi, Dezhi Feng, and Subir Biswas*
*Electrical and Computer Engineering, Michigan State University, East Lansing, MI*

**Abstract:** *This paper presents a deep-learning based traffic classification method for identifying multiple streaming video sources at the same time within an encrypted tunnel. The work defines a novel feature inspired by Natural Language Processing (NLP) that allows existing NLP techniques to help the traffic classification. The feature extraction method is described, and a large dataset containing video streaming and web traffic is created to verify its effectiveness. Results are obtained by applying several NLP methods to show that the proposed method performs well on both binary and multilabel traffic classification problems. We also show the ability to achieve zero-shot learning with the proposed method.*

## I. INTRODUCTION

### *A. Motivation*

The rapid growth of the Internet has helped to popularize various data-intensive applications. For example, the top 2 most heavily accessed video streaming sites, Netflix and YouTube, constitute roughly 50% of the North American Internet traffic. As the popularity of video streaming sites grows, their usage also becomes significant in private enterprises, where watching certain videos may be undesirable. For example, employers may not want their employees to watch certain videos containing political, sexual, or violence related messages. Often the administrator of an enterprise network may want to block users from watching videos in order to conserve bandwidth and/or to maintain enterprise productivity. At the same time, an administrator may wish to allow users to access some of the video streaming sites due to other business reasons. As such, the rise of video streaming applications poses an interesting challenge for network traffic management.

Enforcing the regulations on network traffic first requires the classification of traffic, which is a problem that draws much interest from both the networking community and the business side. It has been thoroughly investigated in the past. Various solutions have been proposed and applied to the problem [1]–[3] Existing solutions fall into two categories: rule-based solutions, and Traffic Analysis (TA) based solutions. They are effective in their respective problem domains, and they have found applications in commercial networking devices.

However, the existing solutions face several challenges. Firstly, the adversaries, who want to avoid detection, upgrade their methods over time, resulting in an arms race between them and the network administrators. Traffic encryption technologies such as Virtual Private Network (VPN) [4] and The Onion Router (Tor) [5] are widely accessible nowadays, and researchers are finding new ways to masquerade traffic [6]. Therefore, the current methods need reevaluation and improvement.

Secondly, the amount and diversity of network traffic have increased drastically during the last decade. Existing solutions might not scale well and need updates to work under the new traffic conditions. Growth in video streaming traffic is arguably the most significant recent change in network traffic, yet there are only a limited number of researches targeting video streaming protocols [7]–[9]. Analysis of heterogeneous traffic (where multiple types of network traffic occur at the same time) is left out of the existing research as well but happens quite often in real-world situations. The targeted traffic type needs to be extended to cover these changes.

In recent years, breakthroughs have been made in machine learning using deep learning methods. The trend has prompted numerous studies to apply these methods to traffic analysis problems. However, despite the perspective and the great interest, the existing works have not realized the full potential of deep learning on traffic classification. Some works [10]–[12] are based on traffic flows (a flow is a series of packets exchanged between 2 specific applications on 2 network nodes) and use hand-crafted features from the traditional works, which are not designed with deep-learning in mind. Others[13], [14] work at the packet level and tend to include the payload as part of the input, which makes them effectively deep-learning based deep packet investigation methods. The requirement of payload data limits their applicability in situations where encryptions make payload inaccessible. In short, existing methods either do not fully take advantage of the method or have limited applicability because of the input. This leaves the definition of novel flow-based features for deep-learning classifiers an open problem that has value.

This research sets out to tackle these challenges by defining a classification method centered on a novel feature that is inspired by deep-learning based Natural Language Processing (NLP) methods. The proposed method uses a simple feature transformation to convert a sequence of packet metadata of traffic flow to a symbolic sequence that looks like natural language. We then develop a deep-learning based classifier that exploits the temporal patterns in this feature. The first advantage of this method over the traditional ones is the extra information encoded in the feature due to its higher temporal resolution, which would help the identification of heterogeneous flows (the traffic flows that are the combination of traffic between different source/destinations and/or different protocols) and the identification of unknown traffic mixes based on already-learned signatures (this ability is called zero-shot learning [15], which helps to reduce training workload). The second advantage of the proposed method is that it relies less on hand-crafted features. The third but not the least is that it enables application of existing NLP techniques, for example, word embedding [16], to TA problems. To the authors' best knowledge, this work marks the first attempt at proposing such a method.

### *B. Application Scenario*

The scenario that the work is based on is shown in Fig. 1. A home/small business customer connects to the Internet through a gateway. The customer establishes an encrypted VPN tunnel from a customer-owned router to an external VPN server through this gateway and then established a small-scale subnetwork behind the router. There might be multiple devices sharing the VPN connection, doing different tasks at the same time. The owner of the gateway wants to analyze the encrypted traffic and understand the website usage of the users of the subnetwork without peeking into the payload. The goal of the engineered solution is to be able to recognize the individual service providers that contributed to the traffic flow. The classifiers are expected to assign one or more labels to a slice of the traffic stream, with a confidence score for each label.

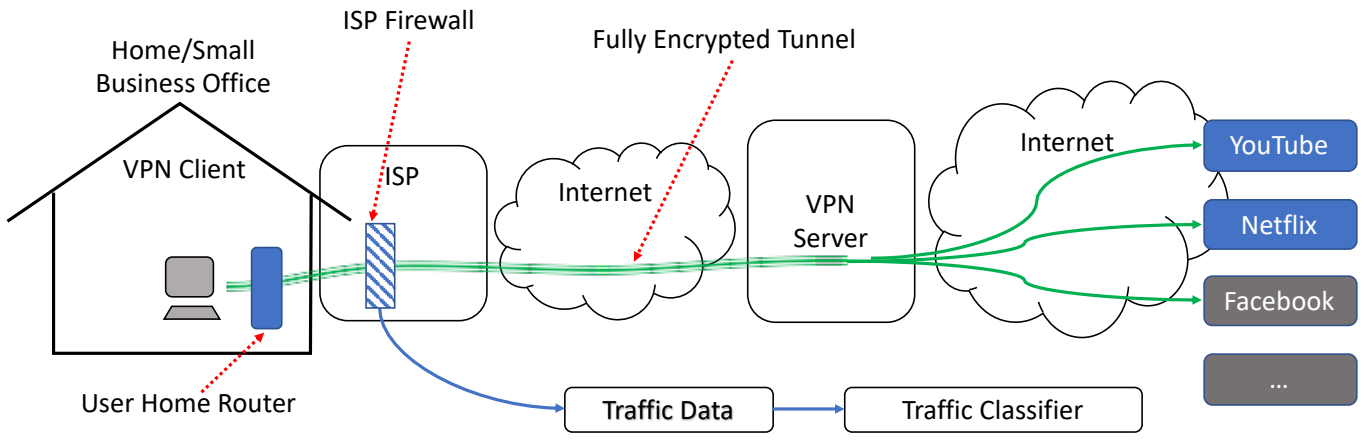

Fig. 1 traffic analysis on encrypted VPN

### C. Main Contribution

In this work we 1) provide a novel method to extract flow-based features from network traffic that is uniquely fit for the application of NLP techniques. 2) Create a multi-label traffic classifier that allows NLP classification methods to be used on network traffic 3) Produce results with state-of-the-art, deep-learning based text classification methods and compare with an established traffic classification method as proof of the effectiveness of the proposed classification method, and 4) Evaluation of the zero-shot learning capability of the feature/classifier.

## II. RELATED WORKS

### A. Traffic Analysis

Traffic Analysis (TA) [17] is a machine-learning based technology to reveal information about encrypted network traffic using data collected from the traffic. The state-of-the-art approach of TA depends on features, data digests created out of the traffic stream, that is used to train machine-learning classifiers to identify the traffic streams and assign labels to them. As the payload is usually not useful in this case, TA works with metadata such as packet direction, packet size, packet count, and timing.

Depending on the application, TA may be called simply Traffic Analysis in the context of traffic classification, or Website Fingerprinting (WFP) [18] in the context of identifying the source of encrypted HTTP traffic. The two sides share the same core technology but focus on different aspects because of the difference in applications. TA focuses on low granularity knowledge, such as protocols or maliciousness, while WFP focuses on high granularity information such as the website, page, and function a user accessed and emphasizes detection through traffic that is cloaked by privacy enhancing protocols such as HTTPS, VPN or Tor protocol. While TA has benign applications, WFP is usually considered a form of adversarial research that challenges the effectiveness of existing protocols in order to provoke improvements.

Numerous works have shown the effectiveness of TA against security enhancing protocols. The work of Hermann et. al [1] is an early piece that established that good recognition can be achieved with a VPN system. Panchenko et. al [2] target the Tor protocol with a combination of features, including traffic amount, packet count, packet size among others, and achieved a significantly higher recognition rate (55%) compared to previous works. This is considered a breakthrough and popularized Tor as a target for further TA based researches, including [19]–[21].

One issue with existing works on TA is that in order to extract statistical features the temporal resolution of the traffic is sacrificed. For example, packet size distribution does not contain information of the time a certain packet size occurs within the sampling window. Even for features that involve time, for example, inter-packet arrival time, the way the feature is extracted eliminates the temporal resolution in the process. This is problematic since the timing of the packets contains potentially useful information on the interaction pattern between the client and the server.

Some researchers acknowledge the problem and try to develop solutions. An early attempt at addressing the issue is reported in [2], where the authors tried to create an n-gram of packet sizes to preserve the packet order information. (An n-gram means an $n$-word long phrase that is treated as an atomic entity when interpreting the text. In this context, it means a sequence of $n$ packets) The mentioning is very brief, only stating that 1) a consecutive sequence of $n$ packet sizes is encoded as one n-gram, and 2) that the resulted feature did not improve the performance. No follow-up works tried this approach again. A possible explanation of this observation is the "curse of dimensionality" of the machine learning classifiers. "Curse of dimensionality" is the phenomenon that increasing the dimensionality of the feature could lead to deteriorated classification performance. The construction of n-grams greatly increases the dimensionality of the feature space, yet the Support Vector Machine-based classifier used in that work performs less well on very high-dimensional features.

The difficulty of exploiting temporal correlation is only part of the problem with feature engineering in TA. Usually, features are defined based on established network metrics, in the hope that what works for humans to understand traffic can be readily transferred into machine learning. However, there is no guarantee of the applicability of a feature to a problem, resulting in a largely trial-and-error based engineering process. The advancement of TA is arguably hindered by the inadequate feature engineering process.

Yet another issue with previous works on TA is that they assume single source/destination in the traffic flow, in other words, all traffic in the tunnel is limited to coming from only one protocol type between one source and one destination. Some researchers already raised concerns on the practicality of WFP because many works are based on this assumption [18]. In reality, the tunneled traffic (e.g., through encrypted VPN or Tor) contains often multiple traffic streams coming from different sources, to different destinations. We call these flows "heterogeneous flows". Due to the abundance of heterogeneous flows, a classifier that can work with them is crucial for the success of TA in the real world.

### *B. Deep Learning*

Deep-learning is a family of biologically inspired machine learning algorithms that feature multiple layers of cascading non-linear representations of data that can be trained. The most popular of deep-learning algorithms is Deep Neural Network (DNN), in which the layers are made of artificial neurons that can be trained through error backpropagation [22]. The learning of multi-layer representations is why DNN is "deep" and is also the reason behind its classification power. Each layer depends on the previous layer(s) and therefore can be considered as a feature extractor. Having multiple layers means there is potential in learning a deep hierarchy of features directly from a high-dimensional representation of the data. In other words, instead of relying on manually engineered features, deep learning algorithms can learn features from raw data. For this reason, deep learning classifiers are commonly trained against densely sampled representations of the data. Images and time series, for example, are often used with minimal preprocessing.

We will focus on the application of deep-learning to NLP problems. The rationale behind the decision is that 1) network traffic flow and text flow are similar in structure, and 2) it enables the classifier to borrow techniques from NLP.

For natural language data, two neural network architectures are used: Convolutional Neural Network (CNN) [23] and Recurrent Neural Network (RNN) [24]. A CNN does a convolution operation on the data with a kernel function that is shared throughout the series, the most significant component is then picked as the input for subsequent layers via max-pooling. An RNN keeps internal states so that it can react to input sequences. The states are updated at each input step by a combination of the past state and the current input. CNN and Variations of RNN, (Especially Long Short-Term Memory, LSTM [25] & Gated Recurrent Unit, GRU [26]), have had great success in NLP[27].

The most notable recent development in DNN based NLP is the Attention-Based Network (ATTN). ATTNs define attention as a neural network (potentially deep) that accepts some form of input and outputs a weight factor that regulates how much weight a part of that input has on the output. ATTN can be chained with other networks to regulate their inputs. Researchers have found that adding attention to a network increases its memory capacity and accuracy. As an example, [28]is regarded as the state-of-the-art text classification result to date.

### *C. Word Embedding*

An efficient word representation is important for NLP. A straightforward method to map words to vectors is one-hot encoding. In one-hot encoding, the *i*th word in a vocabulary that contains *N* words is converted to an *N*-element long vector, with only the *i*th element being 1 while all others being 0. One-hot encoding is simple to implement, but scales poorly to larger vocabularies and does not maintain information on the correlation between words. Word embedding [16] is developed to reduce dimensionality while maintaining the correlation. A Word Embedding is a unique projection from a vocabulary of N words onto N dense vectors in an M-dimensional space (usually M≤N). The embedding ensures that related words (those that appear close to each other in the original sequence) also have vector representations that are similar to each other. An embedding can be trained from a dataset separately from the classifier, and then reused for different tasks to save time. Widely used methods to train embeddings include word2vec [29] and Global Vectors for Word Representation (GLOVE) [30]. Word embedding is an example of the NLP techniques that TA can benefit from.

### *D. Deep Learning based Traffic Analysis*

The introduction of deep learning can be a solution to the problems discussed in section II.A. The features used in deep-learning are much higher in dimensionality, and technologies like Word Embedding can handle high dimensional data with little manual feature engineering. The increased resolution also means that it might be possible to identify individual streams within a heterogeneous flow. There is potential in using the deep-learning technique to solve the feature engineering problem and the heterogeneous flow problem.

Researchers have taken on the challenge of applying the deep-learning technique to traffic analysis problems before. The works can be divided into 3 groups regarding how the features are constructed. The first group contains researches that use purely flow-based, statistical features, like the ones used in traditional TA studies. They are a natural progression from using traditional ML technology but fails to exploit the unique capability of DNNs. Such studies include [12], [31]. The second group is the researches that focus on analyzing the packets. The features used in this group are commonly directly based on the packet payload, and therefore can be regarded as a natural extension of deep-packet investigation with deep learning. Deep-packet is a famous example here [13]. Researches in this group are less applicable to the scenario that this work aims to tackle since analysis of the packet payload is ruled out by the presence of the encrypted tunnel. The third group is where new flow-based features are developed to take advantage of the feature selection capability of DNNs. There are few works in this group. In [11] the researchers extracted a time-series of statistical features from traffic flow. Each point in the series represent 3 seconds of traffic flow, and the resulting time series is used to train a multi-layer LSTM network to classify the traffic streams between the 2 protocols (BitTorrent protocol and plain HTTP protocol) tunneled through an encrypted proxy. The experimental results show that the new features could yield 92% accuracy, which rivals the state-of-the-art non-DNN based methods.

## III. LANGUAGE-LIKE TRAFFIC FEATURE

In this section, we discuss a language-like feature for encoding network traffic. The goal of this feature is 1) achieve high data reduction ratio before the traffic reaches the classifier. Achieving this goal helps reduce the network load when the classifier is eventually deployed to a real-world networking system. 2) produce a language-like feature that existing NLP techniques can process. And 3) Achieve high classification accuracy.

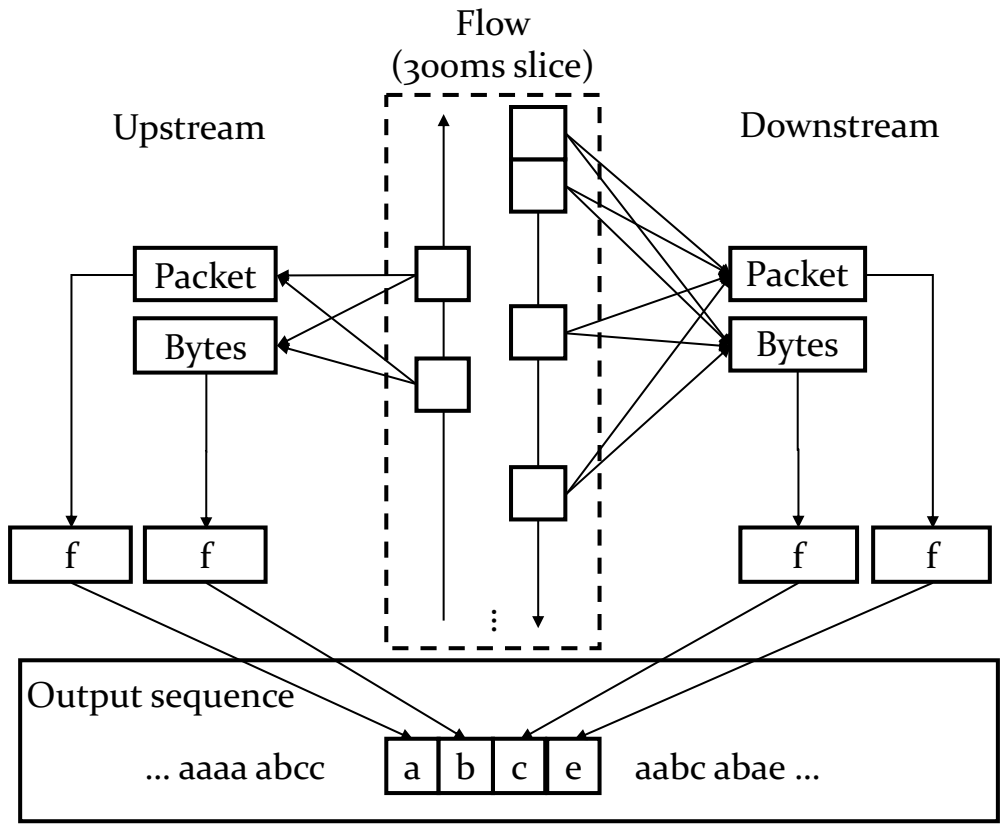

Fig. 2 Extracting Language-Like Traffic Feature

To achieve the goals, we define the feature as depicted in Fig. 2. A 1-minutes-long bidirectional traffic stream is split into 200 non-overlapping windows each 300ms long. Within each window, the sum of upstream/downstream packet counts is taken as well as the upstream/downstream byte counts. Each value is mapped onto the English alphabet using the formula defined in Eqn. 1.

$$f(V) = \lfloor log_{10}(V) \rfloor \qquad \text{Eqn. 1}$$

We use a base-10 log-scale to convert an input value to an index. For a 100Mbps link, which is used in the data collection, the highest possible letter is "g" (the range between $10^7$ and $10^8$). Th reason for a log-scale is that it makes the error tolerance in-scale with the absolute values. It also helps to reduce the number of symbols in the feature. The 4 letters are concatenated to form a single symbol ("word") in the output sequence.

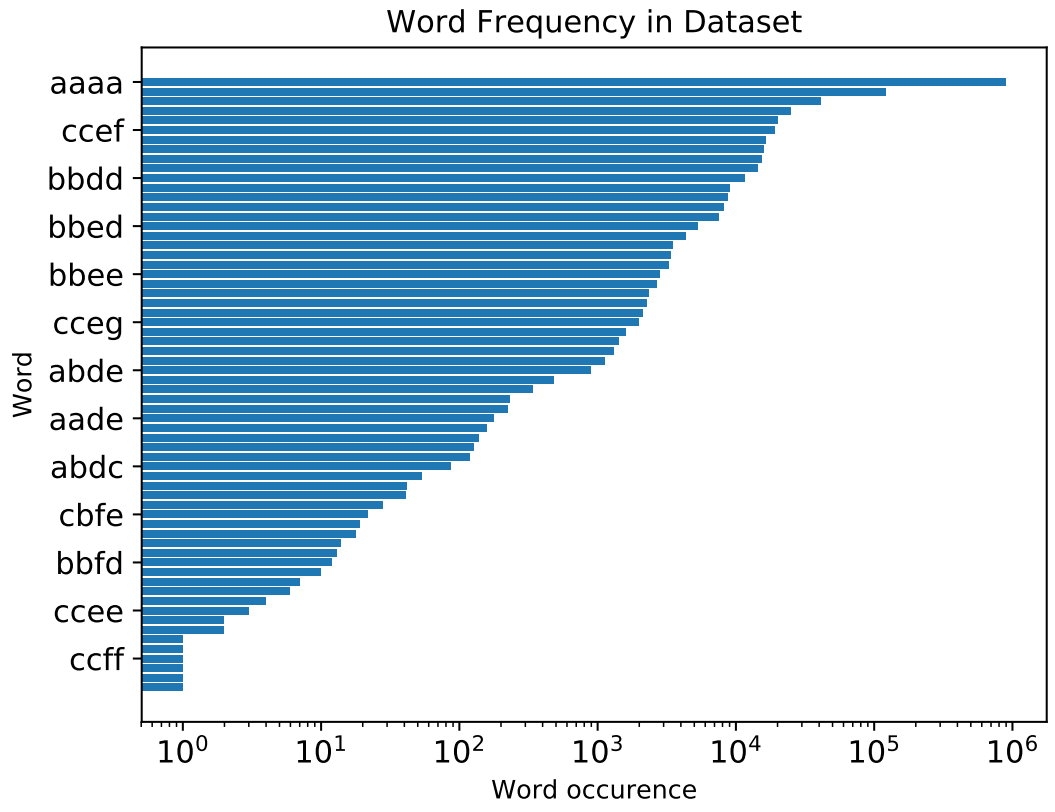

Fig. 3 Word Frequency of Traffic Feature

After the feature extraction, every sample is turned into a sequence of 200 words. Fig. 3 shows the word frequency distribution of the dataset (the making of the dataset will be discussed in section V). The distribution is sorted by popularity, and every 1 in 5 words is labeled on the vertical axis. We can observe that 1) the resultant feature has a vocabulary of only 68 words, a much smaller number than the number of possible words. 2) The word frequency follows an exponential distribution. Each word is about 83.83% as popular as the previous one. A significant anomaly happens at *aaaa*. The overrepresentation of *aaaa* is because it is the word for when the channel is idle and therefore happens more often in the data.

## IV. NEURAL NETWORK ARCHITECTURE

We verify the effectiveness of the feature with the DNN-based text classification methods listed in Table 1. We also choose the traffic classifier by Cruz et. al [11] for our dataset as a comparison.

We define 2 types of classification problems in this work. The first type is a binary classification problem: given unknown traffic, classify it as either containing traffic from a certain video provider or not. The second type is the multi-label problem: given unknown traffic, identify the probability of it containing traffic from each of the video providers in a list. In both cases, the class labels are video streaming providers, but the outputs are categorical in case of the binary problem, and binary in case of the multi-label problem. We use *binary_crossentropy* loss function to train the models in both cases. The performance is measured by *categorical_accuracy* in the binary case and by Receiver Operating Characteristic (ROC) area in the multilabel case.

Word2vec embedding is used across all NLP methods. The embedding is a part of the networks and will be trained alongside them.

Table 1 NLP Processing Methods Used

| Method | Architecture | Optimizer |
|---|---|---|
| **HAN** | GRU+Attention | Adam |
| **Yoon Kim** | CNN | Adam |
| **Mark Berger** | GRU | RMSProp |

The first architecture used is based on Hierarchical Attention Network (HAN) proposed by Zhichao Wang, et. al.[28].

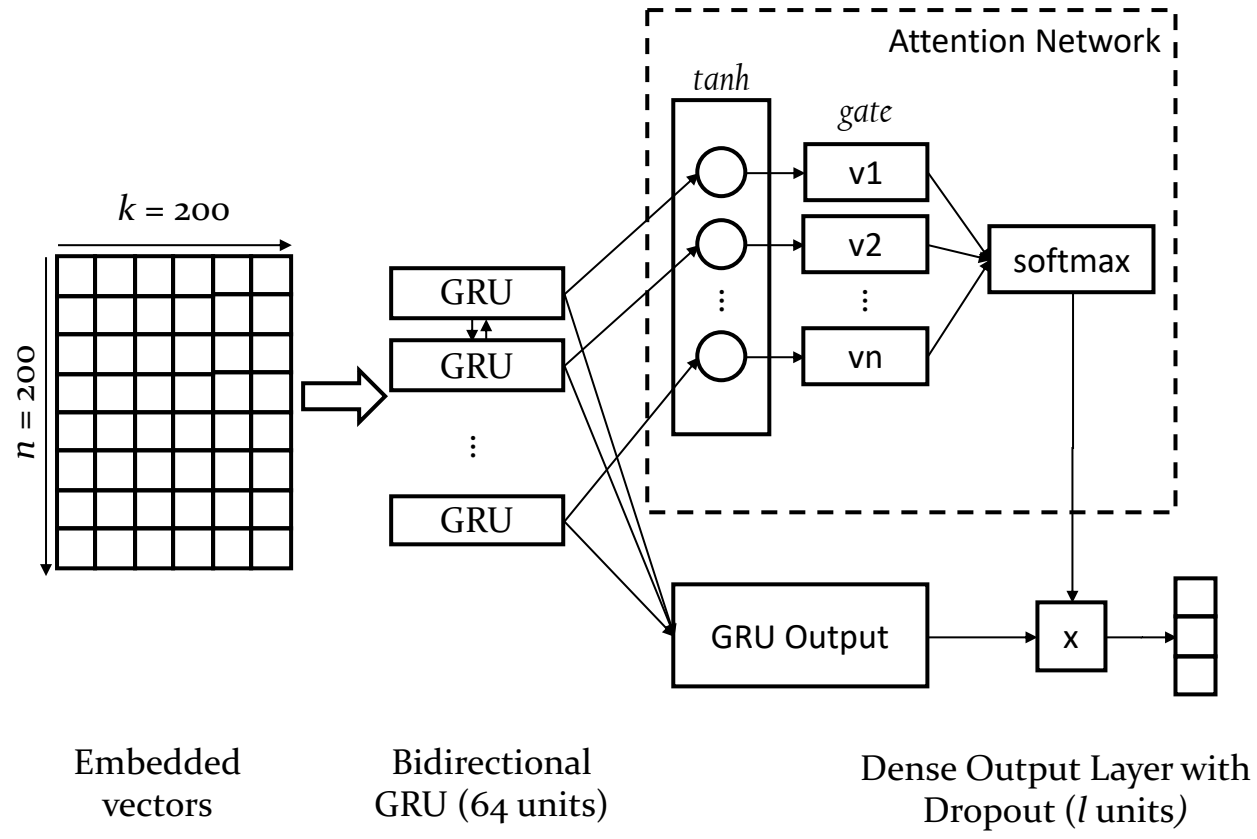

Fig. 4 Hierarchical Attention Network

In this network (Fig. 4), the output is a weighted sum of the outputs of a bidirectional GRU layer, and the weights are from the output of an ATTN. This ATTN is equivalent to the word-level attention network in the original work. We do not need a document-level ATTN as the original work does since there is no higher-level structure in our dataset.

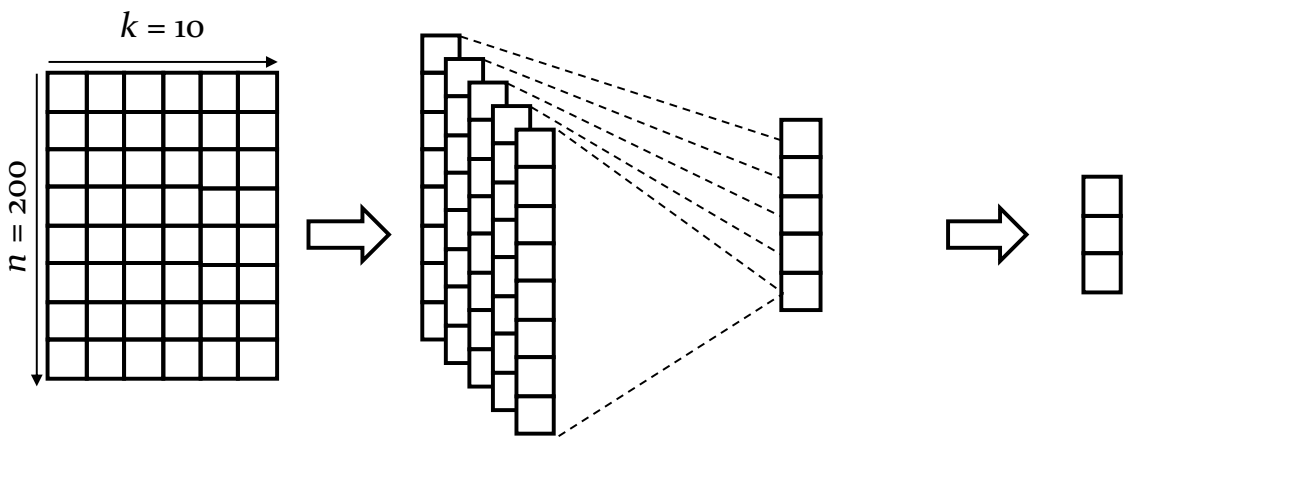

Fig. 5 Network Architecture by Yoon Kim

A 200-dimensional Word2Vec embedding is used as inputs to this network and all other network models. The output is a dense sigmoid layer of labels, with the labels either in categorical format or binary format.

The second method used in this work is proposed by Yoon Kim [32]. This network (depicted in Fig. 5) has multiple 1D-CNN layers with ReLU activation, each having a different kernel shape. According to this method, a layer with a kernel shape *n* is defined for *n*-grams in input data. We use kernels with lengths from 1 to 5 to focus on unigrams and up to 5-grams. Each 1D-CNN layer contains 256 units. The input is 200 10-dimensional Word2Vec embedded vectors. The CNN layers are flattened to a dense vector with max-pooling-over-time. The output layer is the same format as HAN.

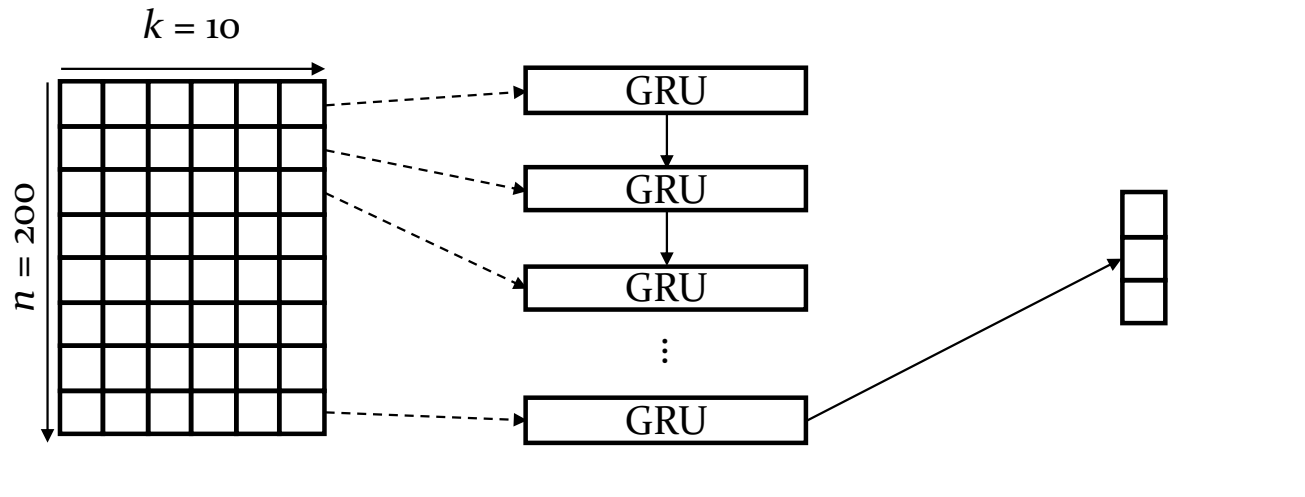

Fig. 6 Network Architecture by Mark Berger

The third method (Fig. 6) used is proposed by Mark Berger [33]. The network accepts the same 10-dimensional embedded vectors as in Fig. 5, and consists of a GRU layer with ReLU activation and a dense output layer that also follows the same format in Fig. 3.

We feature the traffic classification network by M. Cruz et. al as a recent example of a DNN network designed for TA. The network accepts a vector of traffic features. Each feature vector is calculated from 3 seconds of traffic and contains many components. To adapt it for our dataset the TCP specific features (for example roundtrip time and TCP flags) are discarded because we do not assume a TCP-based tunneling protocol. The remaining 60 feature values are listed in Table 2.

Table 2 Features by M. Cruz et. al

| Field | Description |
|---|---|
| **Packet count 1** | Bidirectional packet counts of the 1st second of traffic |
| **Packet count 2** | Same as above for the 2nd second |
| **Packet count 3** | Same as above for the 3rd second |
| **Byte count 1** | Bidirectional byte counts of the 1st second of traffic |
| **Byte count 2** | Same as above for the 2nd second |
| **Byte count 3** | Same as above for the 3rd second |
| **Bidirectional packet size stats over a 3-second interval & Bidirectional inter-arrival interval stats over 3-second interval** | Each stat contains the following values:<br>- maximum value<br>- minimum value<br>- mean value<br>- median value<br>- variation<br>- extreme outlier (>2σ) count<br>- mild outlier (>σ) count<br>- Shannon entropy<br>- 2,3,4 and 5-permutation entropy |

The network (Fig. 7) contains a dense input layer with ReLU activation, then a chain of 6 LSTM layers with various hidden state sizes that have tanh activation, and another dense layer with ReLU activation before the sigmoid output layer. The LSTM layers (except for the last one) output the entire output sequence to allow them to be chained. The original network uses a binary output format and uses *binary_crossentropy* loss function to optimize for the binary classification problem. We change the output layer to use categorical output (in binary problems only) to match the other networks in this work.

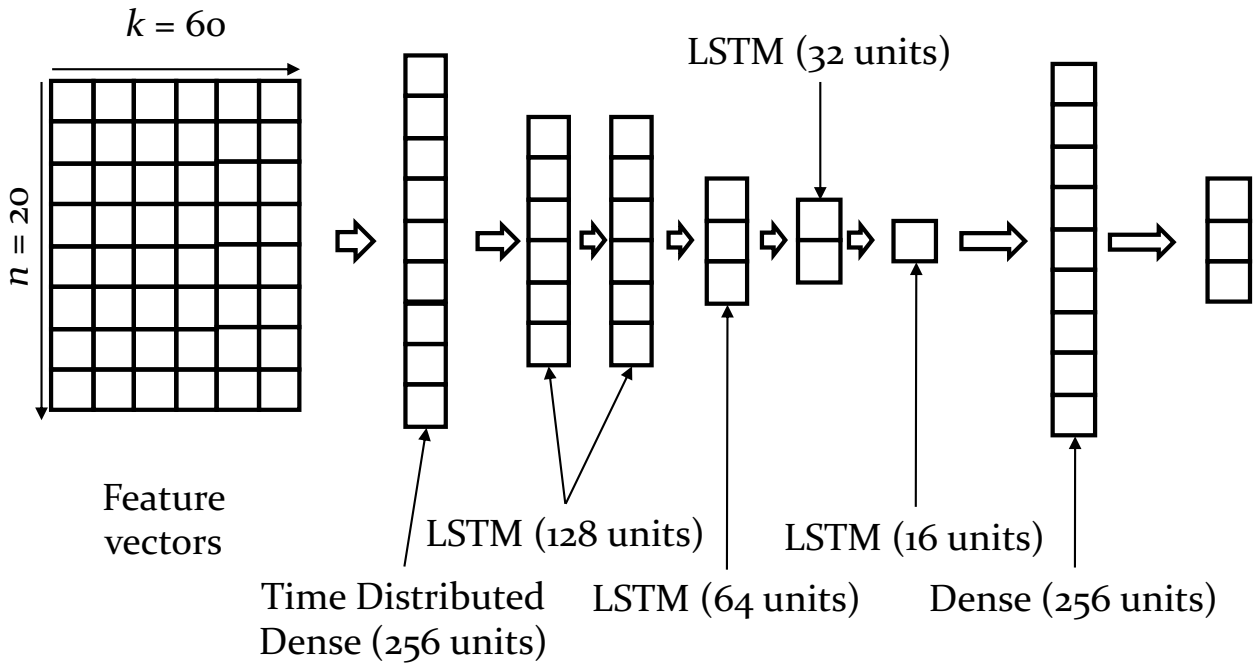

Fig. 7 Network Architecture by M. Cruz et. al.

## V. EXPERIMENTAL SETUP

We construct the network shown in Fig. 8 for the experiments. A Wi-Fi access point runs an OpenVPN[34] client which tunnels to an OpenVPN server. The client and the server talk in the SSL-over-UDP protocol. The tunnel is monitored by a recording router. A brief sample of the recording that the router produces is shown in Table 3 (note that the actual timestamps are in epoch time format, so they can be matched to the ground truth later). Up to 4 clients can connect to the Internet through this tunnel to do different activities. The clients are randomly assigned to watch a video (various length) from one of the six video streaming sites listed in Table 4, to access a website (1min/access), or to stay idle for reference. We reserve one client for accessing a web page every minute to create noise in the dataset. Any sample could contain 0 to 3 video streaming traffic flow and some web traffic flow. The tunnel is also used for day-to-day Internet access when data collection is idle, and the traffic from the casual uses are recorded to create a referential "background" data.

Table 3 Traffic Data Format

| t(ms) | size(byte) | src ip | dst ip | sport | dport |
|---|---|---|---|---|---|
| 0.000 | 1514 | 54.192.39.46 | 192.168.0.95 | 443 | 59666 |
| 0.060 | 1514 | 54.192.39.46 | 192.168.0.95 | 443 | 59666 |
| 0.180 | 1514 | 54.192.39.46 | 192.168.0.95 | 443 | 59666 |
| 0.590 | 66 | 192.168.0.95 | 54.192.39.46 | 59666 | 443 |
| … | ... | … | … | … | … |
| 2.490 | 1514 | 54.192.39.46 | 192.168.0.95 | 443 | 59666 |

The data collection continued for about a month, during which time about 19k 1-minute long samples are collected. The amount of collected data is shown in Table 5. The accessed video streaming sites are recorded and later used to assign labels to the samples. Each sample might be assigned 1-4 labels depending on the client activities at the time of recording. 4-label samples are rare in the dataset (28 samples), the reason is that they only happen between the transitions of videos. These samples are then processed with the proposed feature extraction method to be classified by the 3 NLP-based methods. Features according to the specifications listed in Table 2 are also extracted to be classified by Cruz. et. al's classification method.

In all cases, we randomly choose 80% of the dataset as training set and 20% as the test set. The training set is balanced when evaluating the binary classification problems but is left as is when evaluating the multi-label classification problems.

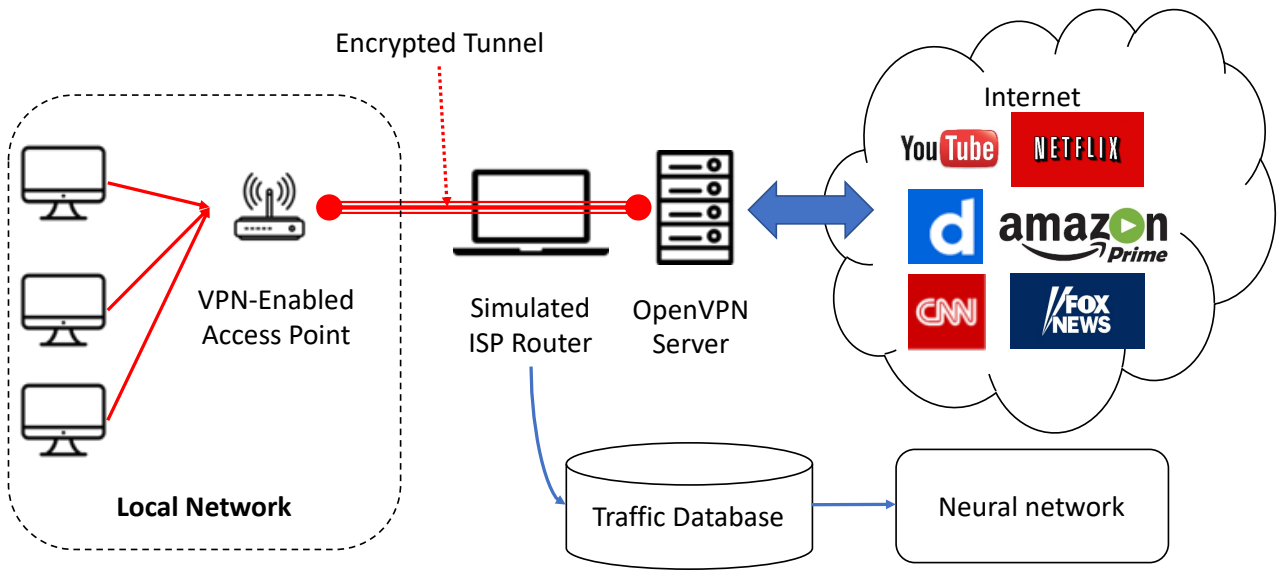

Fig. 8 Experimental Setup

Table 4 Classes based on video streaming sources

| Video Source | Number of videos in the playlist |
|---|---|
| Amazon | 2 |
| CNN News | 1 (autoplays other videos) |
| Fox News | 1 (autoplays other videos) |
| DailyMotion | 1 (autoplays other videos) |
| Netflix | 2 |
| YouTube | 2 (autoplays other videos) |

Table 5 Sample counts by No. of Clients

| All | 1 Client | 2 Clients | 3 Clients | 4 Clients |
|---|---|---|---|---|
| **19321** | 8703 | 9128 | 1402 | 28 |

We will study the following 3 aspects of the performance. 1) The accuracy of the method when solving a binary classification problem. 2) Comparison when solving a multi-label classification problem compared with the binary classification case, and 3) the performance of the testing samples that bear a combination of labels that are absent in the training set. This problem is called the "knockout problem" from here on. Being able to identify such samples is to say the network can do Zero-shot Learning. It is a valuable trait because this ability reduces the training overhead of the network greatly by allowing the network to be trained on an incomplete subset of all possible label combinations.

## VI. EXPERIMENTAL RESULTS

### *A. Binary Classification*

The accuracies of all binary classification experiments are shown in Fig. 9. The result shows that the newly defined feature is effective. The accuracy in binary problems is defined as follows: if the activation level of the neuron corresponding to the label (0 or 1) is higher than the activation level of the other output neuron, it is considered a hit. The accuracies of the binary classifiers vary from 86.3% to 95.7% depending on the video streaming site being focused on. Cruz et. al performs erratically: its accuracy is 65.3% on Netflix and 64.4% on Amazon, but its highest accuracy reaches 92.3% (on Fox). The newly defined feature performs either comparable or significantly better than the Cruz. et. al. method in all cases.

Table 6 Training Performance

| Method | HAN | Y.Kim | M.Berger | Cruz. et.al. |
|---|---|---|---|---|
| **Epoch Time** | 17s | 35s | 54s | 14s |
| **#Epochs** | 30 | 50 | 50 | 50 |

Table 6**Error! Reference source not found.** shows the time needed for each network to complete one training epoch and the number of epochs needed for convergence. The training/testing is done on a setup that consists of TensorFlow 1.11 running on a GeForce 750Ti GPU with 4GB of video memory. While Cruz. et. al.'s method is the fastest to complete an epoch, HAN is the fastest to converge among the methods, with good classification performance achieved at 30 epochs while other methods take about 50 epochs to reach good performance.

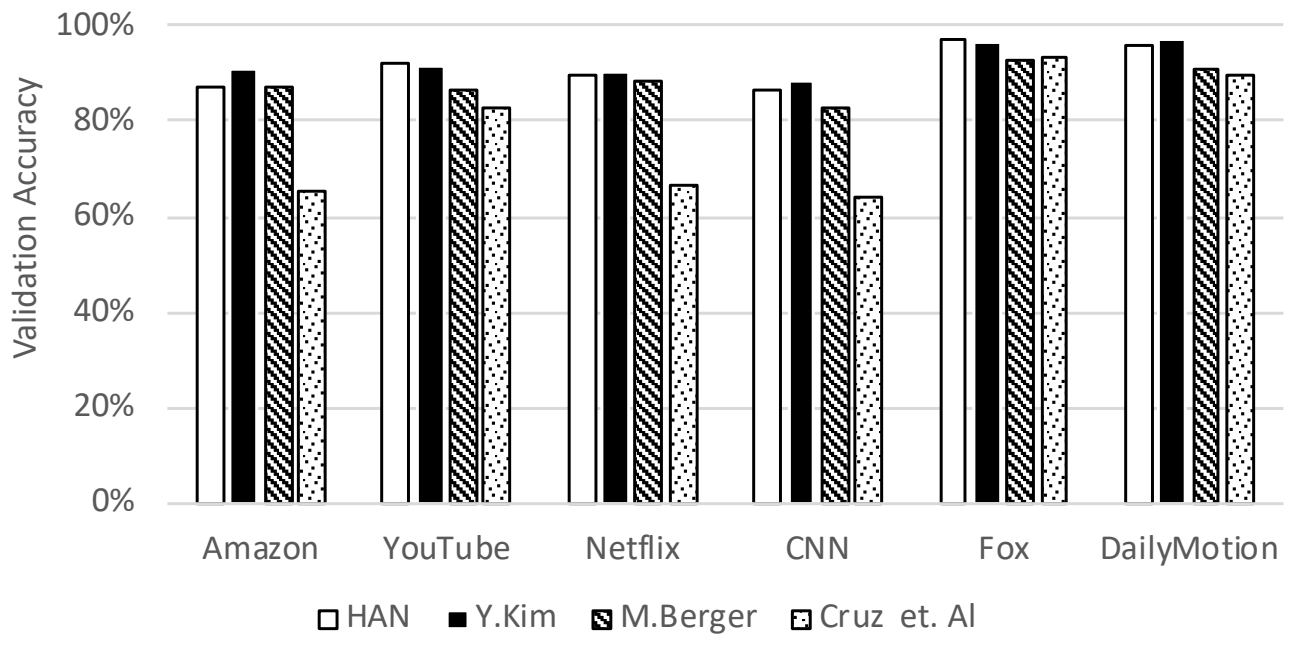

Fig. 9 Binary Classifier Accuracy

Effect of Word Embedding: We highlight the fact that the embedding size used in HAN method is set to $k$=200, while there are only 68 unique words in the dataset. The choice is based on observations featured in Fig. 10, which show that the oversized embedding improves the accuracy (from 88.4% when embedding size=10 to 92.1% when $k$=200). The observation is sensible since in machine learning, mapping the features into a higher-dimensional space in order to find better separations between classes is an established practice. Increasing the embedding size further has the opposite effect. Other methods also benefit from a larger embedding size albeit to a lesser degree. For example, Y. Kim's method gains another 1.3% accuracy on YouTube when its embedding size is increased to 200. A large embedding size, however, costs more memory, which is why the embedding sizes for the other methods are set to 10. These observations hint at potentially interesting uses of embedding in TA applications.

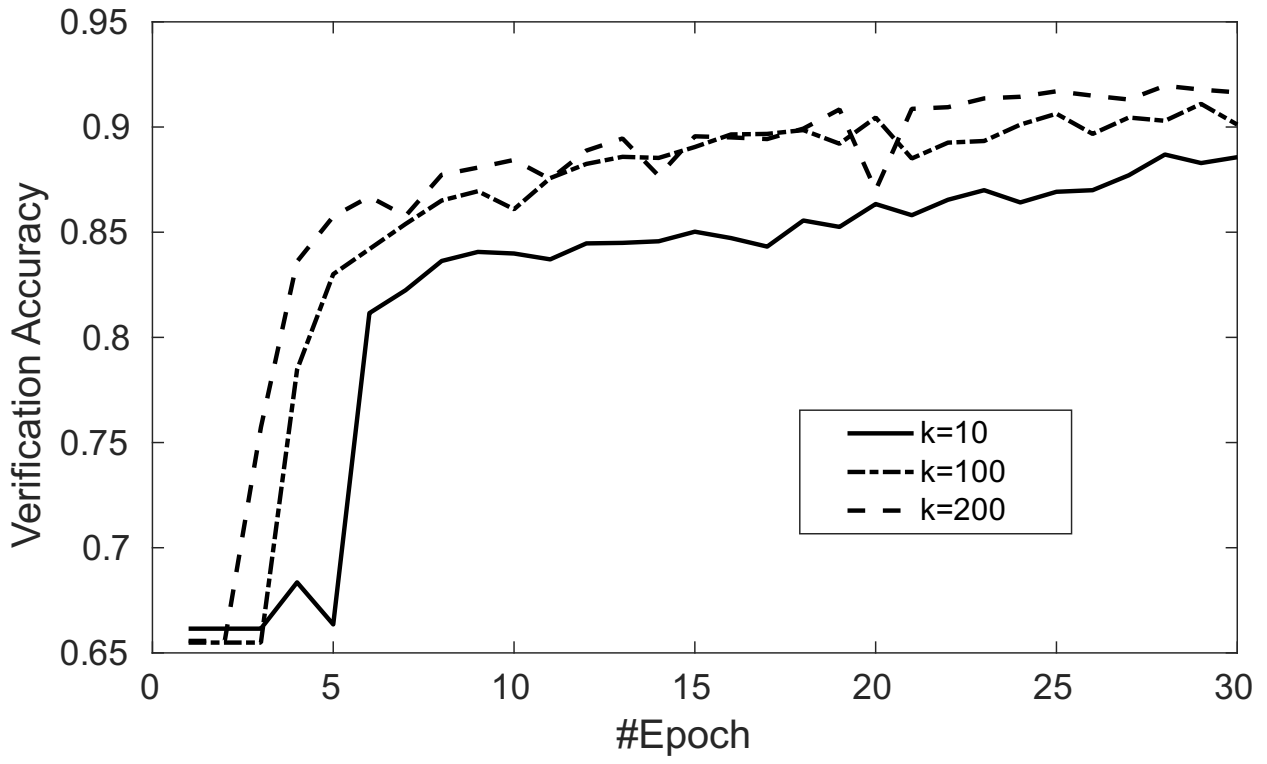

Fig. 10 The Effect of Embedding Size on HAN-based Binary Classifier (YouTube)

### B. Multilabel Classification

To evaluate multilabel classification performance we obtain the activation level at the output stage for all of the methods and convert them into ROC curves. The AUCs would indicate the overall performance. We let the classifiers to classify the 6 video streaming sites together.

For comparison, the AUCs of Y. Kim's binary classifiers are provided in Table 7. We choose them since Y. Kim's classifier performs the best in our binary experiments. These numbers provide the performance reference for the multilabel classifiers.

Table 7 AUC of Binary Classifiers using Y.Kim Method

| Site | Amazon | YouTube | Netflix | CNN | Fox | DailyMotion |
|---|---|---|---|---|---|---|
| AUC | 0.984 | 0.941 | 0.883 | 0.887 | 0.984 | 0.937 |

The ROC curves and their respective AUCs of the multi-label classifiers can be seen in Fig. 11. The horizontal axis in these graphs is the false positive rate, and the vertical axis is the true positive rate. The AUCs can be read from the legends.

Y. Kim's method again performs well across all classes, with HAN following closely. M.Berger's method performs moderately on Amazon and YouTube while showing declining performance on other classes. Cruz et. al.'s method only performs well enough in one class (Fox). The AUCs of Y. Kim method does not change much from the binary AUC values listed in Table 7, proving that the proposed feature is effective in multi-label classification.

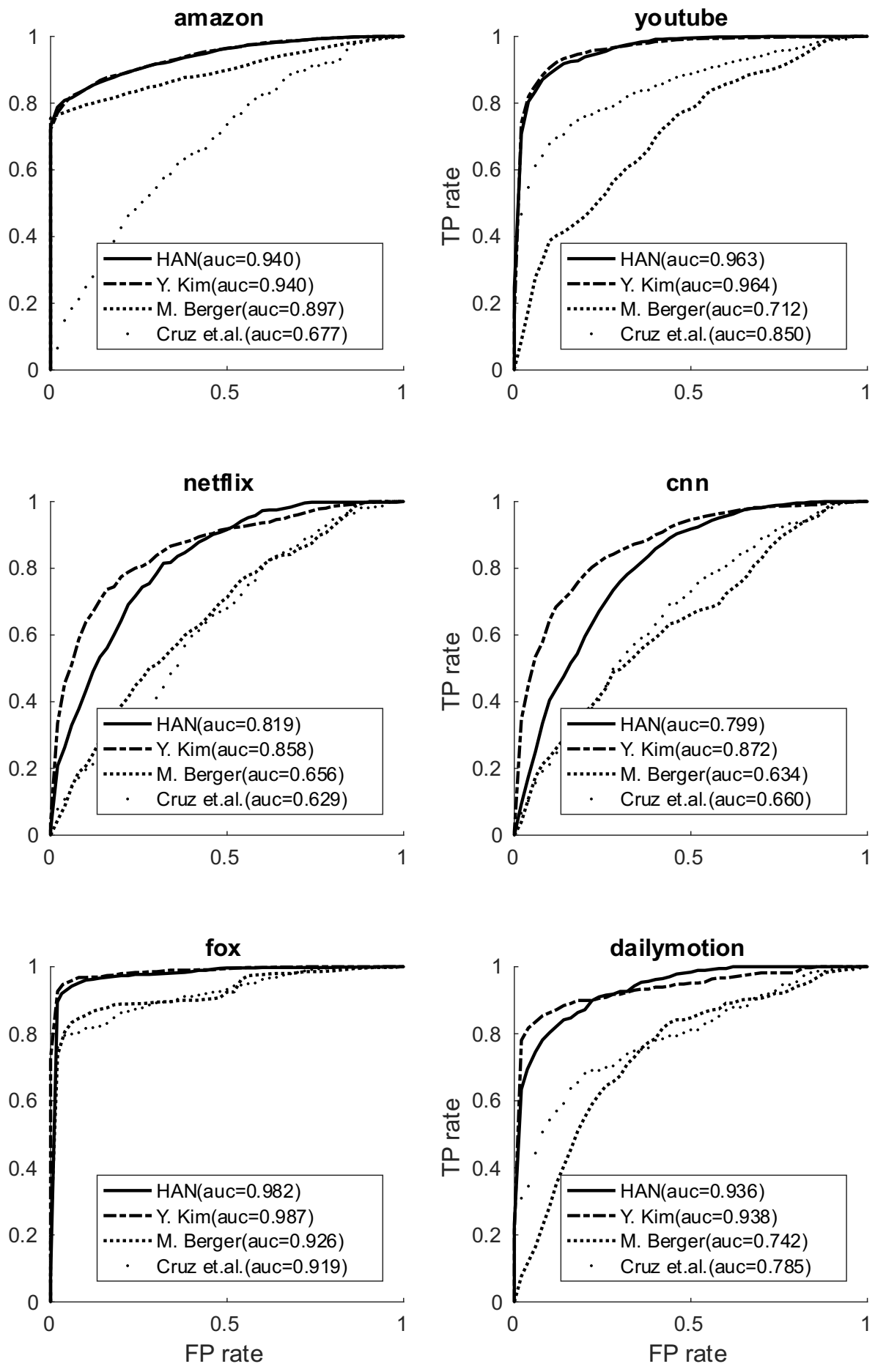

Fig. 11 Multi-label Classifier ROC

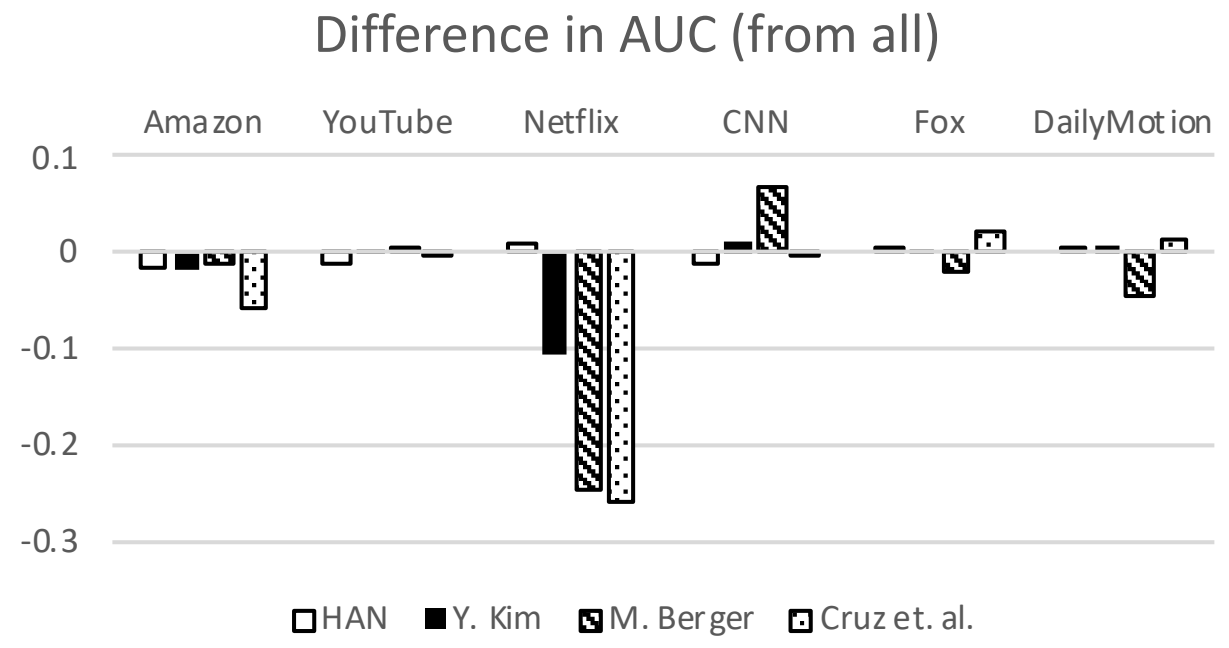

Fig. 12 Multi-label Classifier AUC (No. of Client ≥ 2)

Fig. 13 shows the ROC curves and AUCs of the same multi-label classifiers evaluated on the subset of samples with 3 or more active clients. These samples are considered challenging because of the traffic mixture. We also observe that the AUCs do not change much from those in Fig. 11 when we limit the evaluation to the subset of samples with 2 or more active clients. The difference in AUCs of this subset from the whole set is provided in Fig. 12 for reference. The only significant difference happens on Netflix samples.

According to Fig. 13, the resilience of the classifiers is challenged on the subset where client number is greater than 2. Y. Kim's method again performs well compared to the overall results suffering only minor drops in AUC. YouTube is a major exception, where the method suffered a drop of AUC from 0.964 to 0.808. It still performs better than other methods in all cases. Cruz et al. method actually improves on Fox and DailyMotion but does not perform well on Amazon, Netflix or CNN. The observations here show the effectiveness of the proposed feature applies to the case of multi-label classification.

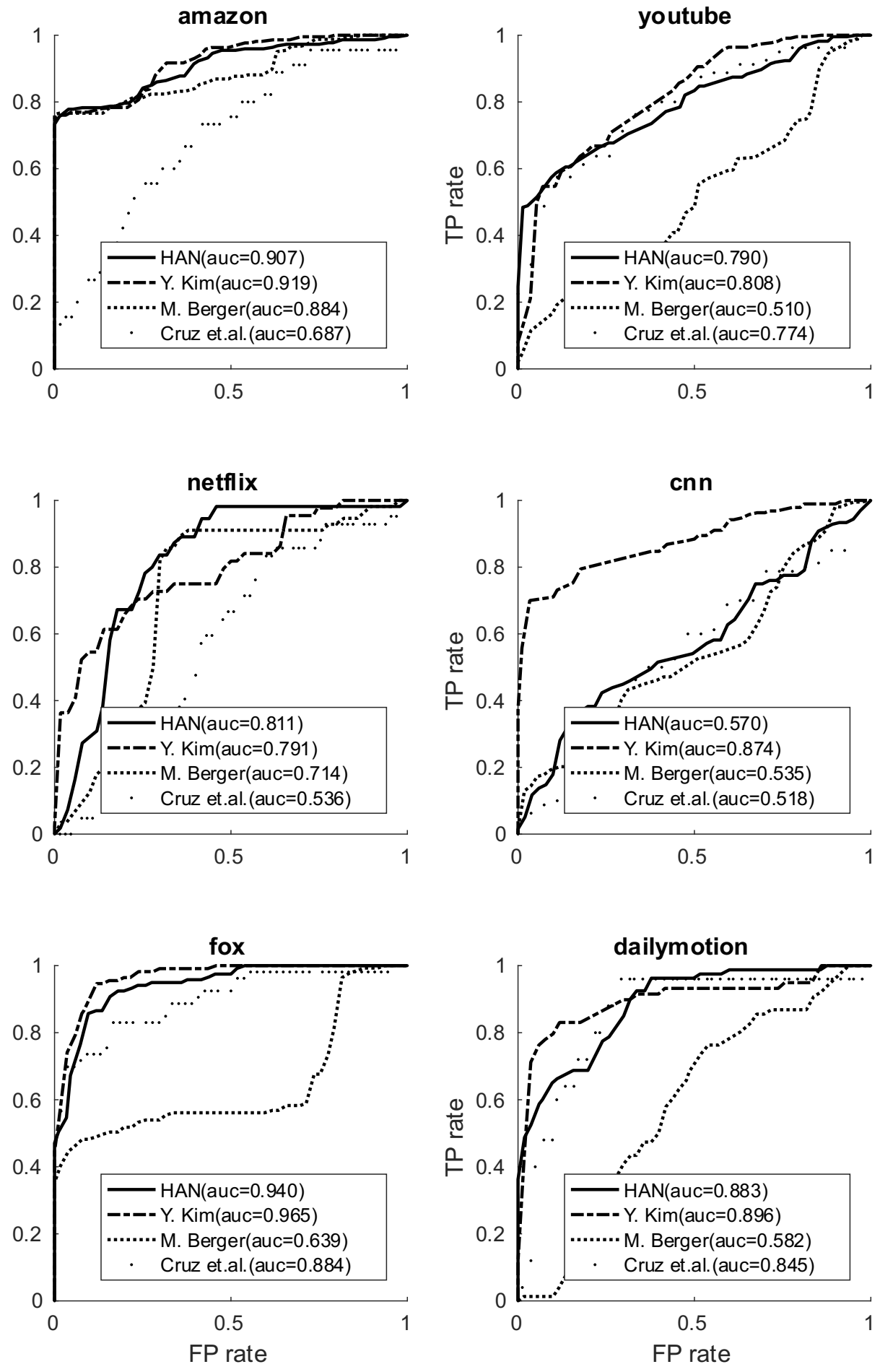

Fig. 13 Multi-label Classifier ROC (No. of Client ≥ 3)

### C. *Knockout Test*

To evaluate the performance of the classifier, we remove samples bearing certain pairs of labels from the training set, one pair at a time, and train HAN and Y.Kim classifier on the same conditions as the multilabel tests using these modified datasets. The trained classifiers are then evaluated on the augmented test sets that contain those previously removed samples. The ideal output is for both output neurons (corresponding to the pair of knocked-out classes) to have high activation levels. The before-and-after comparisons of two pairs, YouTube-Amazon on the left and Amazon-DailyMotion on the right, are shown in Fig. 14. The 1$^{st}$ row is the result before the knockout for reference, the 2$^{nd}$ row is the result of Y. Kim's method after the knockout and the 3$^{rd}$ row is the result of HAN after the knockout. The observations here are generalizable to other cases.

Before the knock-out, the network can be trained to drive both neurons to high activation levels. But while the network can still identify the sites after the removal, it becomes weaker at identifying both at once. For many samples, the network is only able to drive one of the two neurons to high activation levels at a time. This effect is more visible with HAN, for no sample can drive the network output to the top-right corner after the knockout.

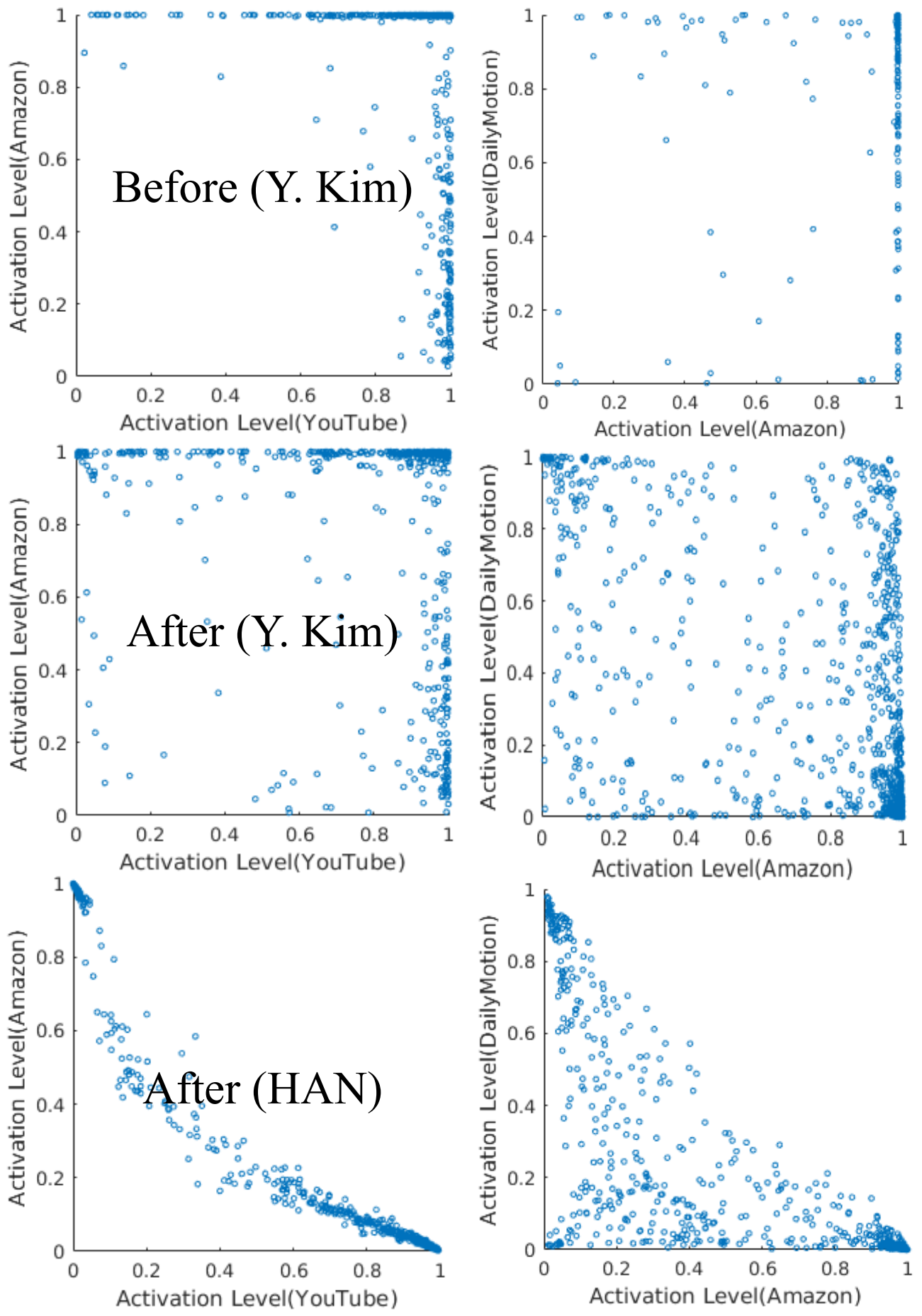

Fig. 14 Selected Results from Knockout Tests

The observation might be a result of the max-pooling mechanism or the attention mechanism. Max-pooling or attention mechanism helps the network focus on the most significant input components (suspected to be the site generating most traffic in that sample) while suppressing other components, resulting in the observed results. The performance might be improved if a different network architecture that can focus on both is used.

## VII. Summary

In this work, we propose a novel classification method for identifying the source of heterogeneous network traffic flow using deep-learning based classifiers. The method consists of a newly defined feature that is inspired by researches in NLP. The feature reduces encrypted network traffic to a concise text-like representation that requires a minimal amount of manual feature engineering. The representation can then be processed by several NLP techniques that potentially will increase classification performance. We combine the feature with a neural network that takes advantage of word embedding in the proposed method.

Using a dedicated experimental setup, we compare the performance between the new method (3 NLP methods are adapted into the classifier) and a traffic classification method that uses hand-crafted features. The results show that the new feature performs better on binary classification problems. We then compare the performance of the methods on multi-label classification problems. Again our method is shown to work well.

Finally, we explore the possibility of classifying traffic containing a combination of labels that are not present in the training dataset ("knockout" problem). The method is shown to be able to identify the predominant component of the traffic.

Future work on our method includes 1) improvement of the feature, for example, include more metadata, 2) explore the effect of parameters on classification performance (sample size, word size etc.), 3) trials on other tunneling protocols and 4) improve its ability to do zero-shot learning.

Another potential direction, although not within the scope of this work, is the potential to break the limitation that the training data and the test data should be collected against the same VPN server/client machine pair. It used to be the case that if the machines are changed then the classification performance suffers, which limits the applicability of TA in real-world problems. Improvement in that direction would make TA methods more powerful.